# Heralded hybrid CV-DV entanglement generation by quantum interference between CV state and DV delocalized photon

Sergey A. Podoshvedov[1] and Nguyen Ba An[2,3]

[1]*Laboratory of Quantum Information Processing and Quantum Computing, Institute of Natural and Exact Sciences, South Ural State University (SUSU), Lenin Av. 76, Chelyabinsk, Russia*
[3]*Institute of Physics, Vietnam Academy of Science and Technology (VAST), 18 Hoang Quoc Viet, Cau Giay, Hanoi, Vietnam*
[4]*Thang Long Institute of Mathematics and Applied Sciences (TIMAS), Thang Long University, Nghiem Xuan Yem, Hoang Mai, Hanoi, Vietnam*

**Abstract**

Hybrid entangled states prove to be necessary for quantum information processing within heterogeneous quantum networks. A method with irreducible number of consumed resources that firmly provides hybrid CV-DV entanglement for any input conditions of the experimental setup is proposed. Namely, a family of CV states is introduced. Each of such CV states is first superimposed on a beam-splitter with a delocalized photon and then detected by a photo-detector behind the beam-splitter. Detection of any photon number heralds generation of a hybrid CV-DV entangled state in the outputs, independent of transmission/reflection coefficients of the beam-splitter and size of the input CV state. Nonclassical properties of the generated state are studied and their entanglement degree in terms of negativity is calculated. There are wide domains of values of input parameters of the experimental setup that can be chosen to make the generated state maximally entangled. The proposed method is also applicable to truncated versions of the input CV states. We also propose a simple method to produce even/odd CV states.

## 1. Introduction

Entanglement, namely, the property of two or more physical systems to be described by one wave function (one state), despite the fact that these physical systems can be at a considerable distance from each other, is the most mysterious fundamental concept in quantum physics [1-4]. Entanglement is the basis for quantum teleportation [5-11], quantum state engineering [12,13] and quantum computing [14-17]. Spontaneous parametric down-conversion (SPDC) has been the most widely used to produce the light entangled states most [18]. The probabilistic nature of the source of entangled states is a major obstacle when scaling to larger systems. Therefore, methods of generation of highly entangled states independent on input conditions is important task. All this has motivated the study of deterministic sources of photonic entangled state [19,20]. For the time being, only specific entangled states have been generated deterministically. Delivering the entanglement, especially in a deterministic fashion, could provide significant facilities for secure long-distance communications and powerful quantum computing. Therefore, the development of faster technologies for creating entanglement, which are experimentally feasible, is an outstanding problem. Achieving greater involvement of practical states in the generation of entanglement is serious challenge which our work is addressed.

Here, we present a light source to firmly generate entanglement between a CV state and a photon for arbitrary initial conditions with an irreducible amount of consumed resources. The hybrid entangled states is type of entanglement formed by objects of various physical nature [21-28]. The potential of such states for quantum information processing is quite high [10,11,29,30]. The entangling operation developed (CV-DV entanglement) is based on quantum interference of CV states with a delocalized photon on a beam splitter with arbitrary parameters with the subsequent registration of any measurement result in one of the modes of the beam splitter. As the CV states, we choose a family of superpositions of displaced Fock states with equal modulus but opposite in sign displacement amplitudes (generalization of Schrödinger cat states). Depending on the parity of the Fock states forming the CV states, they are divided into even and odd. We also propose an approach to the generation of such even/odd CV states using the routinely used in practice single mode squeezed vacuum state which greatly increases practical utility of the method. The obtained degree of entanglement (in terms of negativity [1,2,31,32]) varies in a wide range but never takes zero values. Large choices of experimental parameters of the source provide maximum entanglement with a high success probability. The method is also applicable to truncated versions of such even/odd CV states. The method can become the basis for distribution of the entanglement between distant points of a quantum network regardless of input conditions.

## 2. Family of even/odd CV states
### 2.1 Definition

When considering generating entangled hybrid states, it is quite common to use optical analog of the even/odd Schrödinger cat states ($SCSs$) [33]

$$|\Omega_\pm^{(0)}\rangle = N_\pm^{(0)}(\beta)(|-\beta\rangle \pm |\beta\rangle), \tag{1}$$

where $|\pm\beta\rangle$ are the coherent states of amplitudes $\pm\beta$ with $\beta$ assumed real and positive ($\beta > 0$) throughout for simplicity and $N_\pm^{(0)}(\beta) = \left(2(1 \pm F(2\beta))\right)^{-1/2}$ with $F(\beta) = exp(-|\beta|^2/2)$ is the normalization factor. The states of the large size $\beta$ are hardly realizable in practice due to the impossibility of implementing a sufficiently strong cubic nonlinearity in such a way that the incipient superposition would not be destroyed by decoherence which damps superposition coherence along state propagation [34]. As a rule, researchers deal with either a truncated version of the $SCSs$, which is a superposition of several first Fock states, or with the CV states approximating $SCSs$ with some fidelity [35]. Generation of the superpositions that could approximate $SCSs$ of an amplitude $\beta = 2$ with the fidelity of 0.99 already presents significant practical difficulties [12].

Despite the technological difficulties in the implementation of the $SCSs$, we expand the class of "similar" states that could be used in entangling operation (EO). The family of the CV states is superposition of the displaced number states ($DNSs$) [35-37] whose displacement amplitudes are equal in magnitude but opposite in sign

$$|\Omega_\pm^{(l)}\rangle = N_\pm^{(l)}(\beta)(|l,-\beta\rangle \pm (-1)^l|l,\beta\rangle). \tag{2}$$

In Eq. (2) the $DNSs$ are defined as $|l,\pm\beta\rangle = D(\pm\beta)|l\rangle$ where $D(\beta) = exp(\beta a^+ - \beta^* a)$ is the unitary displacement operator with displacement amplitudes $\beta$, $a$ ($a^+$) is the photon annihilation (creation) operator and $|l\rangle$ is the Fock state containing $l$ photons (see more details in Appendix A). The normalization factor $N_\pm^{(l)}(\beta)$ is given by

$$N_\pm^{(l)}(\beta) = \left(2\left(1 \pm (-1)^l F(2\beta)c_l^{(l)}(2\beta)\right)\right)^{-1/2}, \tag{3}$$

with the coefficients $c_l^{(l)}(\beta)$ defined by Eq. (A1) in Appendix A. In the case of $l = 0$, we have $SCSs$ (1). By analogy with $SCSs$, we name the states in Eq. (2) as superposition of displaced $l$-photon states ($SDlPSs$) of amplitude $\beta$. For example, in the case of $l = 1$, we deal with a superposition of displaced single photon states ($SDSPSs$).

Depending on the parity of the Fock states forming the superpositions, the CV states in Eq. (2) can be divided into even and odd. Indeed, if we use the decomposition of the $DNSs$ in the Fock basis (A1) and the relation between the coefficients $c_n^{(l)}(\beta)$ and $c_n^{(l)}(-\beta)$ specified in Eq. (A3), then we can rewrite the states in Eq. (2) as

$$|\Omega_\pm^{(l)}\rangle = (-1)^l N_\pm^{(l)}(\beta) F(\beta) \sum_{n=0}^{\infty} c_n^{(l)}(\beta)((-1)^n \pm 1)|n\rangle. \tag{4}$$

It follows from Eq. (4) that regardless of value $l$ the states $|\Omega_+^{(l)}(\beta)\rangle$ are nonzero only for $n = 2m$, i.e.,

$$|\Omega_+^{(l)}\rangle = 2(-1)^l N_+^{(l)}(\beta) F(\beta) \sum_{m=0}^{\infty} c_{2m}^{(l)}(\beta)|2m\rangle, \tag{5}$$

while only terms with $n = 2m + 1$ contribute to the states $|\Omega_-^{(l)}(\beta)\rangle$, i.e.,

$$|\Omega_-^{(l)}\rangle = 2(-1)^{l+1} N_-^{(l)}(\beta) F(\beta) \sum_{m=0}^{\infty} c_{2m+1}^{(l)}(\beta)|2m+1\rangle. \tag{6}$$

That is, the states $|\Omega_+^{(l)}\rangle$ consist only of even Fock states, while the states $|\Omega_-^{(l)}\rangle$ involve only odd Fock states. Hence, we call $|\Omega_+^{(l)}\rangle$ even CV states, while $|\Omega_-^{(l)}\rangle$ odd ones, independent of the value of $l$. Because the states $|\Omega_+^{(k)}\rangle$ and $|\Omega_-^{(m)}\rangle$ have different parities, they are orthogonal to each other:

$$\langle \Omega_+^{(k)} | \Omega_-^{(m)} \rangle = 0. \tag{7}$$

As for states of the same parity, they are not mutually orthogonal

$$\langle \Omega_\pm^{(k)} | \Omega_\pm^{(m)} \rangle = 2 N_\pm^{(k)}(\beta) N_\pm^{(m)}(\beta) \left( \delta_{km} \pm (-1)^m F(2\beta) c_k^{(m)}(2\beta) \right), \tag{8}$$

where $\delta_{km}$ is Kronecker's delta symbol. But since this scalar product contains an exponential factor $F(2\beta)$, then the value of the scalar product decreases rather quickly with increasing $\beta$. Therefore, the states can be considered orthogonal in the case of sufficiently large values $\beta$.

Despite the fact that the generation of introduced states (4,5) can still present significant difficulties, the introduction of new CV states expands the possibilities for generation of "similar" superpositions and manipulation with them. Indeed, instead of spending efforts to generate the $SCSs$, one can try to produce one of the state from the set that may be a more successful event. Even more, one may use truncated versions of the $SDlPSs$ that could approximate them with a sufficiently high fidelity for certain values of the amplitude $\beta$ as is usually done in the case of implementation of the $SCSs$. But instead of two truncated versions for even/odd $SCSs$, one can deal with a lot of similar finite superpositions that can be used in optical quantum information processing, e.g.,

$$|\Omega_+^{(l)}\rangle \cong N_+^{(l)} \sum_{m=0}^{l} c_{2m}^{(l)}(\beta)|2m\rangle, \tag{9}$$

$$|\Omega_-^{(l)}\rangle \cong N_-^{(l)} \sum_{m=0}^{l} c_{2m+1}^{(l)}(\beta)|2m+1\rangle, \tag{10}$$

where $N_\pm^{(l)}$ are the corresponding normalization factors. This approximation is possible for certain values of $\beta$ with some fidelity. In these cases, we can talk about the use of even/odd multi-photon states of light.

The choice of the source CV states can be more significant. For example, as even source states, one can choose a finite superposition composed exclusively of a number of even CV states (5) with different values of $l$ like

$$\left|\Omega_+^{(01\ldots l)}\right\rangle = N_+^{(01\ldots l)} \sum_{k=0}^{l} b_+^{(k)} \left|\Omega_+^{(k)}\right\rangle, \quad (11)$$

where $\{b_+^{(k)}; k = 0,1,\ldots,l\}$ are the expansion coefficients and $N_+^{(01\ldots l)}$ is the normalization factor heeding nonorthogonality of the states $\left\{\left|\Omega_+^{(k)}\right\rangle; k = 0,1,\ldots,l\right\}$ between each other. Likewise, as odd source states, one can choose a finite superposition composed exclusively of a number of odd CV states (6) with different values of $l$ like

$$\left|\Omega_-^{(01\ldots l)}\right\rangle = N_-^{(01\ldots l)} \sum_{k=0}^{l} b_-^{(k)} \left|\Omega_-^{(k)}\right\rangle, \quad (12)$$

where $\{b_-^{(k)}; k = 0,1,\ldots,l\}$ are the expansion coefficients and $N_-^{(01\ldots l)}$ is the corresponding normalization factor. Using the states (11,12) can only enhance capabilities for implementation of the EO between "similar" CV states and delocalized photon compared to the case of exclusively $SCSs$ (1). In fact, we need exclusively either even or odd CV states, therefore the amplitudes for either even or odd Fock states in Eqs. (11,12) can take on any values, which greatly facilitates the possibilities of quantum engineering of the CV states. On the other hand, these states in their properties may resemble $SCSs$. It is also worth noting that one can also use truncated versions of the states (11,12) like those presented in Eqs. (9,10), providing that the implementation of the even/odd multi-photon states state does not cause serious technological difficulties. All this testifies in favor of practical feasibility either of the $SDlPSs$ or their truncated versions and use of them for EO. A practical method for generating even/odd CV states is presented in Appendix E.

**2.2 Nonclassical properties**

In the previous section, we noted the fact that the introduced CV states can resemble $SCSs$ in their properties, which, given that these states are also applicable for the EO, does the method universal for realization of the hybrid entanglement from even/odd CV states and delocalized photon. A state of light can be described by the Wigner function which is a kind of quasiprobability distribution. A state whose Wigner function takes some negative values is referred to as nonclassical.

In Fig. 1, we show the dependences of the Wigner function $W$ for three types of states: even/odd $SCSs$ $\left|\Omega_\pm^{(0)}\right\rangle$ (two plots in the top row), even/odd $SDSPSs$ $\left|\Omega_\pm^{(1)}\right\rangle$ (two plots in the middle row) and even/odd states $\left|\Omega_\pm^{(01)}\right\rangle = N_\pm^{(01)}\left(\left|\Omega_\pm^{(0)}\right\rangle + \left|\Omega_\pm^{(1)}\right\rangle\right)$ (two plots in the bottom row) as functions of the quadrature components $x_1$ and $x_2$. For all these plots, the value $\beta = 2$ is taken. As can be seen from Fig. 1, all the three types of the above-mentioned states have regions on the $x_1, x_2$ plane in which the Wigner function takes negative values $W < 0$, which transparently manifest their nonclassicality. Notably, the number of "negative" regions of the Wigner function for the states $\left|\Omega_\pm^{(l)}\right\rangle$ increases with increasing $l$. Concerning the number of "negative" regions of the Wigner function for the case of states $\left|\Omega_\pm^{(01\ldots l)}\right\rangle$, it may be considerably reduced compared to the case of states $\left|\Omega_\pm^{(l)}\right\rangle$ due to interference between states with different $l$. In general, the plots show that the $SDlPSs$ may exhibit rather similar non-classical properties.

Another simple yet quite typical indicator of nonclassicality is the Fano factor [38] which is responsible for the statistics of photocounts and determined by $F = \langle(\Delta N)^2\rangle/\langle N\rangle$ [38], with $N = a^+a$ the photon number operator, $\langle(\Delta N)^2\rangle = \langle N^2\rangle - \langle N\rangle^2$ the photon number variance and $\langle N\rangle$ the averaged photon number. The number of photocounts at the detector output is periodically counted over a certain fixed small sampling time interval. This number fluctuates from experiment to experiment. Repeating the counting process many times gives a set of numbers from which one can obtain the complete probabilistic characteristics of the discrete random number of photocounts. The coherent state, which is the most classical state, has $F = 1$ and its photon number obeys the Poisson distribution. If a state has $F \neq 1$ then its number distribution deviates from the Poisson one. Namely, $F > 1$ corresponds to a super-Poisson (i.e., broader-than-Poisson) photon distribution, but $F < 1$ implies a sub-Poisson (i.e., narrower-than- Poisson) distribution and the associated state is nonclassical. Figure 3 shows how the Fano factor $F$ depends on the size $\beta$ of the states $|\Omega_{\pm}^{(0)}\rangle$, $|\Omega_{\pm}^{(1)}\rangle$ and $|\Omega_{\pm}^{(01)}\rangle$. As can be seen from the plots in Fig. 3, the Fano factor of the states $|\Omega_{-}^{(0)}\rangle$, $|\Omega_{-}^{(1)}\rangle$ and $|\Omega_{\pm}^{(01)}\rangle$ is less than 1 for small values of the amplitude $\beta$ which indicates the manifestation of nonclassical properties of these states.

## 3. Superimposing CV states with delocalized photon

Now, we are interested in proposing schemes to generate optical entangled light which hybridizes macro- and micro-states using the $SDlPSs$ $|\Omega_{\pm}^{(l)}\rangle$ introduced in section 2 as sources. For that purpose we also need an additional delocalized photon in the state
$$|\varphi\rangle_{23} = a_0|01\rangle_{23} + a_1|10\rangle_{23}, \qquad (13)$$
with $a_0, a_1 \neq 0$ and $|a_0|^2 + |a_1|^2 = 1$. Clearly, the photon in the state (13) is delocalizing in the sense that it occupies simultaneously two different spatial modes, modes 2 and 3. The state (13) can easily be prepared by inputting a single photon to a beam-splitter ($BS$) with transmittance $|a_0|^2$ and reflectance $|a_1|^2$. The source state $|\Omega_{\pm}^{(l)}\rangle$ is superimposed with mode 2 of the delocalized photon on a $BS$ which is described by the following unitary matrix $BS = \begin{bmatrix} t & -r \\ r & t \end{bmatrix}$, where $t$ and $r = \sqrt{1-t^2}$ are the real transmission and reflection coefficients, as shown in Fig. 4.

Behind the $BS$, the number of photons in mode 2 is recorded. It follows from the calculations in the Appendixes (B,C), for whatever the number $n$ of recorded photons the output modes 1 and 3 are heralded to be in the state
$$\left|\Delta_{\pm,n}^{(l)}\right\rangle_{13} = \mathfrak{N}_{\pm,n}^{(l)} \left(a_0\left|\Psi_{\pm,n}^{(l)}\right\rangle_1 |1\rangle_3 + a_1 B_{\pm,n}^{(l)}\left|\Phi_{\pm,n}^{(l)}\right\rangle_1 |0\rangle_3\right), \qquad (14)$$
which is a CV-DV hybrid entanglement because the states of mode 1 are CV states while those in mode 3 are DV ones. The technique of the EO based on interfering the CV states with the delocalized photon with subsequent performing photon number resolving ($PNR$) detection in auxiliary mode may resemble photon catalysis of the optical state [39]. Note that near-unity efficient $PNR$ detection is now experimentally available [40]. In Eq. (14) the normalization factors $\mathfrak{N}_{\pm,n}^{(l)} = \left(|a_0|^2 + |a_1|^2\left|B_{\pm,n}^{(l)}\right|^2\right)^{-1/2}$ depend on the amplitudes $B_{\pm,n}^{(l)}$ which are determined in the Appendixes (B,C). Interestingly, the explicit expressions of $\left|\Psi_{\pm,n}^{(l)}\right\rangle$ and $\left|\Phi_{\pm,n}^{(l)}\right\rangle$ in Eq. (14) are subject to both the parity subindices " $\pm$ " of the CV source states $|\Omega_{\pm}^{(l)}\rangle$ and the parity of the recorded photon number $n$ of mode 2. Namely, if the source states are $|\Omega_{\pm}^{(l)}\rangle$ and $n$ is even ($i.e., n = 2m$) or odd ($i.e., n = 2m + 1$), then the CV states of the output mode 1 appear to be
$$\left|\Psi_{\pm,2m}^{(l)}\right\rangle_1 = L_{\pm,2m}^{(l)} \sum_{p=0}^{l} x_{\pm,2m,p}^{(l)} \left|\Omega_{\pm}^{(p)}\right\rangle_1, \qquad (15)$$

$$|\Phi_{\pm,2m}^{(l)}\rangle_1 = K_{\pm,2m}^{(l)} \sum_{p=0}^{l+1} y_{\pm,2m,p}^{(l)} |\Omega_{\mp}^{(p)}\rangle_1, \qquad (16)$$

$$|\Psi_{\pm,2m+1}^{(l)}\rangle_1 = L_{\pm,2m+1}^{(l)} \sum_{p=0}^{l} x_{\pm,2m+1,p}^{(l)} |\Omega_{\mp}^{(p)}\rangle_1, \qquad (17)$$

$$|\Phi_{\pm,2m+1}^{(l)}\rangle_1 = K_{\pm,2m+1}^{(l)} \sum_{p=0}^{l+1} y_{\pm,2m+1,p}^{(l)} |\Omega_{\pm}^{(p)}\rangle_1, \qquad (18)$$

where $L_{\pm,2m}^{(l)}$, $K_{\pm,2m}^{(l)}$, $L_{\pm,2m+1}^{(l)}$, $K_{\pm,2m+1}^{(l)}$ are the normalization factors and $x_{\pm,2m,p}^{(l)}$, $y_{\pm,2m,p}^{(l)}$, $x_{\pm,2m+1,p}^{(l)}$, $y_{\pm,2m+1,p}^{(l)}$ the expansion coefficients, whose analytical expressions are derived in the Appendixes (B,C). Since $|\Omega_+^{(p)}\rangle$ are even CV states and $|\Omega_-^{(p)}\rangle$ are odd CV states for any values of $p$, it is transparent that each of the output CV states on the LHS of Eqs. (14-18) can only be either even or odd. Table 1 summaries all the possible dependences of parity of the output CV states $|\Psi\rangle_1$ and $|\Phi\rangle_1$ in Eq. (14) on parity of the source CV state $|\Omega\rangle_1$ in Eq. (2) and parity of the photon number $n$ recorded in the output mode 2.

| Pariity of initial CV state $|\Omega\rangle_1$ | even | | odd | |
|---|---|---|---|---|
| Parity of recorded photon number n | even | odd | even | odd |
| Parity of output CV state $|\Psi\rangle_1$ | even | odd | odd | even |
| Parity of output CV state $|\Phi\rangle_1$ | odd | even | even | odd |

**Table 1.** The parity of the output CV states $|\Psi\rangle_1$ and $|\Phi\rangle_1$ in Eq. (14) in dependency on parity of initial CV state $|\Omega\rangle_1$ in Eq. (2) and parity of the detected photon number $n$ in the output mode 2.

The CV-DV hybrid entangled state $|\Delta_{\pm,n}^{(l)}\rangle_{13}$ between the output modes 1 and 3 in Eq. (14) is generated in our scheme with a finite probability $P_{\pm,n}^{(l)}$. In Appendixes (B,C), we present derivation of analytical formulae of $P_{\pm,n}^{(l)}$ for both $n = 2m$ and $n = 2m + 1$.

A more general case involves use of the CV states in Eqs. (11,12) which are superposition of the CV states in Eq. (2). Due to the linearity of the $BS$ operation, one can write

$$BS_{12}\left(|\Omega_{\pm}^{(01...l)}\rangle_1 |\varphi\rangle_{23}\right) = N_{\pm}^{(01...l)} \sum_{k=0}^{l} b_{\pm}^{(k)} BS_{12}\left(|\Omega_{\pm}^{(k)}\rangle_1 |\varphi\rangle_{23}\right). \qquad (19)$$

Each term $BS_{12}\left(|\Omega_{\pm}^{(k)}\rangle_1 |\varphi\rangle_{23}\right)$ contributes to the generated entanglement. All the contributions are summed up to yield the overall conditional states which have the form given by Eq. (14), where the parity of the generated states $|\Psi\rangle_1$ and $|\Phi\rangle_1$ follows from Table 1 and also depends on the parity of the input state as well as the parity of the measurement outcomes. But, nevertheless, amplitudes of the CV states have a rather complex form shown in Appendix D.

It can also be shown that if one uses even/odd multi-photon states (9,10) instead of the CV, the result will be the same. Indeed, an entangled state is also generated in the case of registration of any measurement outcome in the second auxiliary mode excluding $2m + 1$ measurement outcome in the case of the input state (9) and $2m+2$ measurement outcome in the case of the input state (10). The conditional states have the same form as in Eqs. (14). The only difference is that the CV states forming the entanglement are replaced by the finite superpositions. All this indicates the broad applicability of the two-photon fusion by interference of the even/odd CV states with delocalized photon.

## 4. Entanglement degree

The generated states (14) exhibit hybrid entanglement between CV states in mode 1 and DV states in mode 3. As can be verified from Eqs. (15-18) as well as from Table 1, for a fixed set of indices $\{l, \pm, n\}$, each of the states $|\Psi_{\pm,n}^{(l)}\rangle_1$ and $|\Phi_{\pm,n}^{(l)}\rangle_1$ always has a certain parity, but their own parities are different, i.e., if $|\Psi_{\pm,n}^{(l)}\rangle_1$ is even then $|\Phi_{\pm,n}^{(l)}\rangle_1$ is odd, and if $|\Psi_{\pm,n}^{(l)}\rangle_1$ is odd then $|\Phi_{\pm,n}^{(l)}\rangle_1$ is even. Hence, the CV states in mode 1 can be treated as living in a two-dimensional Hilbert space $\mathcal{H}_1$ with two possible orthogonal basis states $\{|even\rangle_1, |odd\rangle_1\}$, where $|even\rangle$ ($|odd\rangle$) implies state that comprises Fock states exclusively containing even (odd) photon numbers. As for the discrete state in mode 3, it lives also in a two-dimensional Hilbert space $\mathcal{H}_3$ with two apparent orthogonal basis states $\{|0\rangle_3, |1\rangle_3\}$. Therefore, Hilbert space of the hybrid entangled states (14) is $\mathcal{H}_{13} = \mathcal{H}_1 \otimes \mathcal{H}_3$ which is four-dimensional with four possible orthogonal basis states $\{|even\rangle_1|0\rangle_3, |odd\rangle_1|0\rangle_3, |even\rangle_1|1\rangle_3, |odd\rangle_1|1\rangle_3\}$.

Since $a_0, a_1 \neq 0$ are assumed, Eq. (14) for the generated states $|\Delta_{\pm,n}^{(l)}\rangle_{13}$ shows disentanglement when the amplitudes $B_{\pm,n}^{(l)}$ vanish. The analytical expressions for $B_{\pm,2m}^{(l)}$ and $B_{\pm,2m+1}^{(l)}$ are derived in Eqs. (B12,B19,C6,D8) and in Eqs. (B13,B20,C11), respectively, which are completely determined by the values of initial experimental parameters. It can be verified that both the amplitudes $B_{\pm,2m}^{(l)}$ and $B_{\pm,2m+1}^{(l)}$ never take zero values, implying that the states $|\Delta_{\pm,n}^{(l)}\rangle_{13}$ generated by our method always possess a finite degree of entanglement. Or, in other words, the generated state is firmly hybrid entangled one.

Entanglement degree of the hybrid entangled states $|\Delta_{\pm,n}^{(l)}\rangle_{13}$ in Eq. (14) can be estimated by using positive partial transpose (PPT) criterion for separability [1,2,31,32]. The negativity $\mathcal{N}$ has all required properties for the entanglement measure. The negativity value ranges from $\mathcal{N}_s = 0$ (separable state) up to $\mathcal{N}_{max} = 1$ (maximally entangled state). One can calculate the negativities $\mathcal{N}_{\pm,2m}^{(l)}$ and $\mathcal{N}_{\pm,2m+1}^{(l)}$ of the states $|\Delta_{\pm,2m}^{(l)}\rangle_{13}$ and $|\Delta_{\pm,2m+1}^{(l)}\rangle_{13}$ (which can be regarded as living in a four-dimensional Hilbert space as aforementioned). The obtained results read

$$\mathcal{N}_{\pm,2m}^{(l)} = \frac{2|a_0||a_1|\left|B_{\pm,2m}^{(l)}\right|}{|a_0|^2+|a_1|^2\left|B_{\pm,2m}^{(l)}\right|^2}, \qquad (20)$$

$$\mathcal{N}_{\pm,2m+1}^{(l)} = \frac{2|a_0||a_1|\left|B_{\pm,2m+1}^{(l)}\right|}{|a_0|^2+|a_1|^2\left|B_{\pm,2m+1}^{(l)}\right|^2}. \qquad (21)$$

As recognized from Eqs. (20,21), the negativities $\mathcal{N}_{\pm,2m}^{(l)}$ and $\mathcal{N}_{\pm,2m+1}^{(l)}$ never vanish (i.e., the conditional states in Eqs. (14) always possess some degree of entanglement) because both $B_{\pm,2m}^{(l)}$ and $B_{\pm,2m+1}^{(l)}$ are nonzero for relevant values of the initial experimental parameters. The maximum value of the negativity is obtained under the condition of either $|a_0| = |a_1| |B^{(l)}_{\pm,2m}|$ or $|a_0| = |a_1| |B^{(l)}_{\pm,2m+1}|$. For example, maximally entangled states can be generated if the balanced delocalized photon $(|a_0| = |a_1| = 1/\sqrt{2})$ is used together with the conditions $|B^{(l)}_{\pm,2m}| = 1$ and $|B^{(l)}_{\pm,2m+1}| = 1$. As can be seen from the analytical expressions for the parameters $B^{(l)}_{\pm,2m}$ in Eqs. (B12,B19,C6,D8) and $B^{(l)}_{\pm,2m+1}$ in Eqs. (B13,B20,C11), the conditions $|B^{(l)}_{\pm,2m}| = 1$ and $|B^{(l)}_{\pm,2m+1}| = 1$ can be met by adjusting the values of initial experimental parameters. Thus, for any source states belonging to family of the states in Eq. (2) the output states in modes 1 and 3 of our scheme in Fig. 5 always appear as CV-DV hybrid entangled states regardless of the number of recorded photon in mode 2. However, the degree of entanglement of the output states is subject to the input experimental parameters as well as to the measurement outcome.

We plot in Fig. 5 the negativities $\mathcal{N}^{(0)}_{+,0}, \mathcal{N}^{(0)}_{+,1}, \mathcal{N}^{(1)}_{+,0}$ and $\mathcal{N}^{(1)}_{+,1}$ (left column) as well as the probabilities $P^{(0)}_{+,0}, P^{(0)}_{+,1}, P^{(1)}_{+,0}$ and $P^{(1)}_{+,1}$ (right column) of successful generation of the hybrid entangled states $|\Delta^{(0)}_{+,0}\rangle_{13}, |\Delta^{(0)}_{+,1}\rangle_{13}, |\Delta^{(1)}_{+,0}\rangle_{13}$ and $|\Delta^{(1)}_{+,1}\rangle_{13}$, respectively, for even source states $|\Omega^{(0)}_+\rangle$ and $|\Omega^{(1)}_+\rangle$ in dependency on $\beta$ and $t$. As can be seen from the figure, there is a fairly large range of values $(\beta, t)$ in which negativity can take on rather large values close to its maximal one $\mathcal{N}_{max} = 1$. Note that the success probabilities can also take rather large values in the given range of experimental parameters. In Fig. 6, we also show the $\mathcal{N}^{(0)}_{-,0}, \mathcal{N}^{(0)}_{-,1}, \mathcal{N}^{(1)}_{-,0}$ and $\mathcal{N}^{(1)}_{-,1}$ (left column) as well as the probabilities $P^{(0)}_{-,0}, P^{(0)}_{-,1}, P^{(1)}_{-,0}$ and $P^{(1)}_{-,1}$ (right column) of successful generation of the hybrid entangled states $|\Delta^{(0)}_{-,0}\rangle_{13}, |\Delta^{(0)}_{-,1}\rangle_{13}, |\Delta^{(1)}_{-,0}\rangle_{13}$ and $|\Delta^{(1)}_{-,1}\rangle_{13}$, respectively, for odd source states $|\Omega^{(0)}_-\rangle$ and $|\Omega^{(1)}_-\rangle$ in dependency on $\beta$ and $t$. They also have areas of parameters $(\beta, t)$ in which the negativity can take values close to $\mathcal{N}_{max} = 1$. All the plots in Fig. 5 and Fig. 6 are constructed for case of the balanced delocalized photon (11) with $a_0 = a_1 = 1/\sqrt{2}$.

The overall conditional states following from (19) can also be described in a four-dimensional Hilbert space regardless of the parity of the input state and parity of the measured photons in mode 2. Thus, one can also use expressions (20,21) for calculating the negativity of the overall conditioned states in the case when the inputs to mode 1 are the CV states $|\Omega^{(01...l)}_\pm\rangle_1$ of Eqs. (11,12). In general, calculating negativity and success probability for an arbitrary number $l$ superposition terms in Eq. (19) is difficult and tedious. In a particular case when the states inputted to mode 1 are $|\Omega^{(01)}_\pm\rangle_1 = N^{(01)}_\pm (|\Omega^{(0)}_\pm\rangle_1 + |\Omega^{(1)}_\pm\rangle_1)$ we calculated the negativities $\mathcal{N}^{(01)}_{+,0}, \mathcal{N}^{(01)}_{+,1}, \mathcal{N}^{(01)}_{-,0}$ and $\mathcal{N}^{(01)}_{-,1}$ as well as the success probabilities $P^{(01)}_{+,0}, P^{(01)}_{+,1}, P^{(01)}_{-,0}$ and $P^{(01)}_{-,1}$ of the output conditional states $|\Delta^{(01)}_{+,0}\rangle_{13}, |\Delta^{(01)}_{+,1}\rangle_{13}, |\Delta^{(01)}_{-,0}\rangle_{13}$ and $|\Delta^{(01)}_{-,1}\rangle_{13}$, respectively. Plots of these calculated quantities are presented in Fig. 7 in dependency on experimental parameters $\beta$ and $t$ in the case of $a_0 = a_1 = 1/\sqrt{2}$. It is interesting to note that a fairly smooth shape is observed for the negativities $\mathcal{N}^{(01)}_{+,0}$ and $\mathcal{N}^{(01)}_{-,0}$, while the shape of surfaces of $\mathcal{N}^{(01)}_{+,1}$ and $\mathcal{N}^{(01)}_{-,1}$ have sharp drops.

Numerical simulations show that the domain of parameter values with which the maximum negativity $\mathcal{N}_{max} = 1$ is observed is very large. Some values of the experimental parameters $(\beta, t)$ that make the negativity maximum are presented in Table 2 for the case of balanced delocalized photon $a_0 = a_1 = 1/\sqrt{2}$. Note that numerical calculations, which we do not present here, show that the maximum entanglement is also observed in the case of an unbalanced delocalized photon $a_0 \neq a_1$ in a large number of cases. As can be seen from the constructed plots in Figs. 5-7 and received data, the values of the experimental parameters can be chosen in such a way that the probability of generating the maximum entanglement can take values close to unity.

| Source state | $\beta$ | $t$ | $n$ | Probability |
|---|---|---|---|---|
| $\left|\Omega_+^{(0)}\right\rangle$ | 0.5 | 0.25 | 0 | 0.939 |
| $\left|\Omega_+^{(0)}\right\rangle$ | 1.4 | 0.65 | 1 | 0.288 |
| $\left|\Omega_+^{(1)}\right\rangle$ | 0.5 | 0.73 | 0 | 0.491 |
| $\left|\Omega_+^{(1)}\right\rangle$ | 0.5 | 0.61 | 1 | 0.301 |
| $\left|\Omega_-^{(0)}\right\rangle$ | 0.5 | 0.79 | 0 | 0.544 |
| $\left|\Omega_-^{(0)}\right\rangle$ | 0.5 | 0.25 | 1 | 0.843 |
| $\left|\Omega_-^{(1)}\right\rangle$ | 0.5 | 0.8 | 0 | 0.523 |
| $\left|\Omega_-^{(1)}\right\rangle$ | 2.1 | 0.96 | 1 | 0.278 |
| $\left|\Omega_+^{(01)}\right\rangle$ | 0.92 | 0.25 | 0 | 0.938 |
| $\left|\Omega_+^{(01)}\right\rangle$ | 1.9 | 0.62 | 1 | 0.291 |
| $\left|\Omega_-^{(01)}\right\rangle$ | 1.34 | 0.8 | 0 | 0.509 |
| $\left|\Omega_-^{(01)}\right\rangle$ | 0.5 | 0.68 | 1 | 0.31 |

**Table 2.** Source states inputted to mode 1, values of the experimental parameters $(\beta, t)$ and the number of recorded photons in mode 2 $(n)$ with which the maximum negativity $\mathcal{N}_{max} = 1$ of the generated hybrid entangled state in Figs. 5-7 is observed. Corresponding success probabilities are also presented.

As noted above, the truncated versions (9,10) for original CV states can also be used as key components of source of hybrid entangled state in Fig. 3. In the case, the conditional states can also be described in four-dimensional Hilbert space as in the case with the input original CV states. The negativity of the entanglement is calculated by the equations (20,21). The only difference is that the original CV states forming the entanglement are replaced by the finite superpositions. Numerical calculations show that the resulting maximum entanglement $\mathcal{N}_{max} = 1$ is also observed in a wide choice of the experimental parameters which indicates the broad applicability of the approach to the source of the entangled hybrid state implementation.

## 5. Conclusion

We offered EO to firmly generate hybrid entangled states between a CV state and a single photon under arbitrary initial conditions. Any $SDlPSs$ can be directly used to generate the conditional entanglement regardless of the initial conditions and measurement outcomes in auxiliary mode. The explanation of the effect can be traced to the example of even $SDlPSs$ and even measurement outcome $2m$. If even number of photons comes from even $SDlPSs$, then heralded state can only comprise even Fock states as even number of photons is detected at auxiliary mode. In other case, if even Fock states of the even $SDlPSs$ are mixed with single photon, the resulting state can only involve odd Fock states in the case of registration of even number of photons in auxiliary mode. Due to indistinguishability of the events, the conditional hybrid entangled state is generated. The same explanation applies to the three remaining cases characterized by the parity of the input and the measurement outcomes. Note that the state of a two-mode squeezed vacuum in the regime of small squeezing amplitude ($r \ll 1$) can also be used for the EO instead of a delocalized photon. Indeed, the output nonnormalised state can be representable as $|00\rangle + \lambda|11\rangle$, where $\lambda$ is a parameter proportional to $r$. This state is entangled with the CV state with help of the same mechanism as is the case with a delocalized photon, where now $a_0 \sim 1$ and $a_1 \sim \lambda$. It results in the entangled state in Eq. (14) with one exception that the following permutation of the states $|0\rangle \rightarrow |1\rangle$ and $|1\rangle \rightarrow |0\rangle$ takes place.

The generated states have certain degree of entanglement characterized by the negativity. Negativity is largely determined by the parameter $B_{2m}^{(l\pm)}, B_{2m+1}^{(l\pm)}$ occurring due to interaction of multiphoton states at the beam splitter. This parameter always takes nonzero values, indicating that hybrid entanglement is always generated under all possible experimental conditions. Large number of the experimental parameters ensures the negativity of the conditional states to takes on maximal value. In addition, the experimental parameters can be selected in such a way to provide a sufficiently high success probability of the state generation with maximal entanglement. The entangled light source is also implemented in the case of truncated versions of the initial CV states. The EO is realized with an irreducible number of linear optics elements which increases the significance of the proposed approach since this can reduce the practical costs associated with EO. In perspective, the source of the hybrid entangled light can be extended to deterministically generate large-scale quantum networks. This can be done through sequential spreading the entanglement between parts of the incipient multipartite state in the same manner.

For EO, we used a family of the CV states being the superposition of the $DNSs$ with equal modulus but different in sign displacement amplitudes. The family of the CV states is a generalization of the well-known $SCSs$ being optical analogue of Schrödinger cat states [33]. As in the case of the $SCSs$, the $SDlPSs$ are divided into even and odd depending on the parity of the Fock states forming a superposition. We constructed the Wigner functions some of the $SDlPSs$ and showed that they have inherent nonclassical properties like regions on phase plane, where the Wigner functions take on negative values. We have also suggested a method for generating even/odd CV states from original single-mode squeezed vacuum state. Taking into account the proposed mechanism for generating CV states using $SMSV$ state, this approach is economical in terms of consumed resources, which nevertheless guarantees the generation of the CV-DV entanglement with large enough degree of entanglement and success probabilities (Figs. 5-7).

## Appendix A. Notes about $DNSs$

Consider the $DNSs$ in the number states basis [35-37]

$$|l, \alpha\rangle \equiv D(\alpha)|l\rangle = F(\alpha) \sum_{n=0}^{\infty} c_n^{(l)}(\alpha) |n\rangle, \tag{A1}$$

where the unitary displacement operator is $D(\beta) = exp(\beta a^+ - \beta^* a)$ with amplitude $\beta$ and $a$ ($a^+$) are bosonic annihilation (creation) operator. The normalization factor is $F(\alpha)$ whose expression was already given in Section 2. The expansion coefficients amplitudes $c_n^{(l)}(\alpha)$ are calculated as

$$c_n^{(l)}(\alpha) = exp(|\alpha|^2/2)\langle n|l, \alpha\rangle, \tag{A2}$$

that provides normalization condition $exp(-|\alpha|^2) \sum_{m=0}^{\infty} c_m^{(l)*}(\alpha) c_m^{(n)}(\alpha) = \delta_{ln}$ for any numbers $l$ and $n$, where $\delta_{ln} = 1$ if $l = n$ and $\delta_{ln} = 0$ if $l \neq n$. As can be shown in [37], the following relation holds

$$c_m^{(n)}(-\alpha) = (-1)^{m-n} c_m^{(n)}(\alpha). \tag{A3}$$

**Appendix B. Superimposing *SCSs* with delocalized photon**

The source states in Eq. (2) with $l = 0$ reduce to the *SCSs* in Eq. (1). Consider interaction of the even *SCSs* $|\Omega_+^{(0)}\rangle$ in mode 1 with a photon delocalized over modes 2 and 3 as in Eq. (13) on a general beam-splitter (i.e., a *BS* with finite transmission (reflection) coefficient $t$ ($r$)). The linearity of the beam-splitter operator implies

$$BS_{12}\left(|\Omega_+^{(0)}\rangle_1 |\varphi\rangle_{23}\right) = N_+^{(0)}(\beta)\left(BS_{12}(|-\beta\rangle_1|\varphi\rangle_{23}) + BS_{12}(|\beta\rangle_1|\varphi\rangle_{23})\right). \tag{B1}$$

For the first term in the parentheses of the RHS of Eq. (B1) we have

$$BS_{12}(|-\beta\rangle_1|\varphi\rangle_{23}) = BS_{12}D_1(-\beta)(|0\rangle_1|\varphi\rangle_{23}) = BS_{12}D_1(-\beta)BS_{12}^+BS_{12}(|0\rangle_1|\varphi\rangle_{23}) =$$
$$D_1(-\beta t)D_2(\beta r)(a_0|00\rangle_{12}|1\rangle_3 + a_1(t|01\rangle_{12} + r|10\rangle_{12})|0\rangle_3) =$$
$$(a_0|0, -\beta t\rangle_1|0, \beta r\rangle_2|1\rangle_3 + a_1(t|0, -\beta t\rangle_1|1, \beta r\rangle_2 + r|1, -\beta t\rangle_1|0, \beta r\rangle_2))|0\rangle_3, \tag{B2}$$

where we embraced by unitarity of the beam splitter operator $BS_{12}BS_{12}^+ = BS_{12}^+BS_{12} = I$ with $I$ being identity operator. The same transformations apply to the second term in the parentheses of the RHS of Eq. (25) that yield

$$BS_{12}(|\beta\rangle_1|\varphi\rangle_{23}) = BS_{12}D_1(\beta)(|0\rangle_1|\varphi\rangle_{23}) = BS_{12}D_1(\beta)BS_{12}^+BS_{12}(|0\rangle_1|\varphi\rangle_{23}) =$$
$$D_1(\beta t)D_1(-\beta r)(a_0|00\rangle_{12}|1\rangle_3 + a_1(t|01\rangle_{12} + r|10\rangle_{12})|0\rangle_3) =$$
$$(a_0|0, \beta t\rangle_1|0, -\beta r\rangle_2|1\rangle_3 + a_1(t|0, \beta t\rangle_1|1, -\beta r\rangle_2 + r|1, \beta t\rangle_1|0, -\beta r\rangle_2))|0\rangle_3. \tag{B3}$$

Using Eqs. (B2) and (B3), one can write the final expression for the RHS of Eq. (B1) as

$$BS_{12}\left(|\Omega_+^{(0)}\rangle_1 |\varphi\rangle_{23}\right) = N_+^{(0)}(\beta)\big(a_0(|0, -\beta t\rangle_1|0, \beta r\rangle_2 + |0, \beta t\rangle_1|0, -\beta r\rangle_2)|1\rangle_3 +$$
$$a_1\big(t(|0, -\beta t\rangle_1|1, \beta r\rangle_2 + |0, \beta t\rangle_1|1, -\beta r\rangle_2) + r(|1, -\beta t\rangle_1|0, \beta r\rangle_2 +$$
$$|1, \beta t\rangle_1|0, -\beta r\rangle_2)\big)|0\rangle_3\big). \tag{B4}$$

Now we can use the decomposition of the displaced states in the Fock basis (A1) taking into account the properties of the matrix elements when changing the sign of the displacement amplitude $\alpha$ to the opposite $\alpha \to -\alpha$ given by Eq. (A3). These bring (B4) to

$$BS_{12}\left(|\Omega_+^{(0)}\rangle_1 |\varphi\rangle_{23}\right) = N_+^{(0)}(\beta)F(\beta r)\sum_{n=0}^{\infty}\big(a_0 c_{0n}(\beta r)(|0, -\beta t\rangle_1 + (-1)^n|0, \beta t\rangle_1)|1\rangle_3 +$$
$$a_1\big(tc_{1n}(\beta r)(|0, -\beta t\rangle_1 + (-1)^{n-1}|0, \beta t\rangle_1) + rc_{0n}(\beta r)(|1, -\beta t\rangle_1 +$$
$$(-1)^n|1, \beta t\rangle_1)\big)|0\rangle_3\big)|n\rangle_2. \tag{B5}$$

Measurement outcomes in mode 2 can be divided into two types depending on the parity of the number $n$ of detected photons: either even $n = 2m$ or odd $n = 2m + 1$. So, if even number of photons $n = 2m$ is registered in mode 2, then the hybrid entangled state in Eq. (14) is generated with the component CV states in Eqs. (15,16) whose expansion coefficients are the following

$$x_{+,2m,0}^{(0)} = 1, \tag{B6}$$

$$y^{(0)}_{+,2m,0} = 1, \tag{B7}$$

$$y^{(0)}_{+,2m,1} = \frac{rc^{(0)}_{2m}(\beta r)N^{(0)}_-(\beta t)}{tc^{(1)}_{2m}(\beta r)N^{(1)}_+(\beta t)}. \tag{B8}$$

In the case of detecting $n = 2m+1$ photons in mode 2, the expansion coefficients of the component CV states in Eqs. (17,18) are

$$x^{(0)}_{+,2m+1,0} = 1, \tag{B9}$$

$$y^{(0)}_{+,2m+1,0} = 1, \tag{B10}$$

$$y^{(0)}_{+,2m+1,1} = \frac{rc^{(0)}_{2m+1}(\beta r)N^{(0)}_+(\beta t)}{tc^{(1)}_{2m+1}(\beta r)N^{(1)}_-(\beta t)}. \tag{B11}$$

The parameters $B^{(0)}_{+,2m}$ and $B^{(0)}_{+,2m+1}$ which to a large extent defines the value of negativity are given by

$$B^{(0)}_{+,2m} = \frac{tc^{(1)}_{2m}(\beta r)N^{(0)}_+(\beta t)}{c^{(0)}_{2m}(\beta r)N^{(0)}_-(\beta t)K^{(0)}_{+,2m}}, \tag{B12}$$

$$B^{(0)}_{+,2m+1} = \frac{tc^{(1)}_{2m+1}(\beta r)N^{(0)}_-(\beta t)}{c^{(0)}_{2m+1}(\beta r)N^{(0)}_+(\beta t)K^{(0)}_{+,2m+1}}. \tag{B13}$$

The corresponding success probabilities to generate the conditional hybrid entangled states are the following

$$P^{(0)}_{+,2m} = \frac{F^2(\beta r)\left|c^{(0)}_{2m}(\beta r)\right|^2 N^{(0)2}_+(\beta)}{N^{(0)2}_+(\beta t)\mathfrak{N}^{(0)2}_{+,2m}}, \tag{B14}$$

$$P^{(0)}_{+,2m+1} = \frac{F^2(\beta r)\left|c^{(0)}_{2m+1}(\beta r)\right|^2 N^{(0)2}_+(\beta)}{N^{(0)2}_-(\beta t)\mathfrak{N}^{(0)2}_{+,2m+1}}, \tag{B15}$$

Similar considerations apply to the odd SCS $|\Omega^{(0)}_-\rangle$ and the resulting component CV states have the following expansion coefficients

$$x^{(0)}_{-,2m,0} = x^{(0)}_{-,2m+1,0} = y^{(0)}_{-,2m,0} = y^{(0)}_{-,2m+1,0} = 1, \tag{B16}$$

$$y^{(0)}_{-,2m,1} = \frac{rc^{(0)}_{2m}(\beta r)N^{(0)}_+(\beta t)}{tc^{(1)}_{2m}(\beta r)N^{(1)}_-(\beta t)}, \tag{B17}$$

$$y^{(0)}_{-,2m+1,1} = \frac{rc^{(0)}_{2m+1}(\beta r)N^{(0)}_-(\beta t)}{tc^{(1)}_{2m+1}(\beta r)N^{(1)}_+(\beta t)}, \tag{B18}$$

while the parameters $B^{(0)}_{-,2m}$ and $B^{(0)}_{-,2m+1}$ become

$$B^{(0)}_{-,2m} = \frac{tc^{(1)}_{2m}(\beta r)N^{(0)}_-(\beta t)}{c^{(0)}_{2m}(\beta r)N^{(0)}_+(\beta t)K^{(0)}_{-,2m}}, \tag{B19}$$

$$B^{(0)}_{-,2m+1} = \frac{tc^{(1)}_{2m+1}(\beta r)N^{(0)}_+(\beta t)}{c^{(0)}_{2m+1}(\beta r)N^{(0)}_-(\beta t)K^{(0)}_{-,2m+1}}. \tag{B20}$$

and the corresponding success probabilities read

$$P^{(0)}_{-,2m} = \frac{F^2(\beta r)\left|c^{(0)}_{2m}(\beta r)\right|^2 N^{(0)2}_-(\beta)}{N^{(0)2}_-(\beta t)\mathfrak{N}^{(0)2}_{-,2m}}, \tag{B21}$$

$$P^{(0)}_{-,2m+1} = \frac{F^2(\beta r)\left|c^{(0)}_{2m+1}(\beta r)\right|^2 N^{(0)2}_-(\beta)}{N^{(0)2}_+(\beta t)\mathfrak{N}^{(0)2}_{-,2m+1}}. \tag{B22}$$

By direct summation, it can be shown that the probabilities sum to one, i.e. $\sum_{m=0}^{\infty}\left(P^{(0)}_{\pm,2m} + P^{(0)}_{\pm,2m+1}\right) = 1$, as should be.

**Appendix C. Superimposing the states (5,6) with delocalized photon**

Now, we are going to consider interaction of the general source states in Eq. (1) with an arbitrary value of $l$ with a delocalized photon on a general $BS$. To do this let us for convenience write down the explicit output from the general $BS$ when the inputs are $l$ photons in mode 1 and the vacuum or a single photon in mode 2, namely

$$BS_{12}(|l\rangle_1|0\rangle_2) = \sum_{k=0}^{l}(-1)^k t^{l-k}r^k\sqrt{\frac{l!}{k!(l-k)!}}|l-k\rangle_1|k\rangle_2 \tag{C1}$$

and

$$BS_{12}(|l\rangle_1|1\rangle_2) = \sqrt{l+1}\,t^l r|l+1\rangle_1|0\rangle_2 +$$
$$\sum_{k=0}^{l}(-1)^k \frac{t^{l-k-1}r^k}{k!}\sqrt{\frac{(k+1)!l!}{(l-k)!}}\left(t^2 - \frac{l-k}{k+1}r^2\right)|l-k\rangle_1|k+1\rangle_2. \tag{C2}$$

These states are the basis for the derivation of the conditional states in Eqs. (14).

As summarized in Table 1, for the general even source state $|\Omega_+^{(l)}\rangle_1$ in Eq. (2) (i.e., $l$ can be any integer including zero), the explicit expression of the generated hybrid entangled state depends on the parity of $n$ (the number of photons in the output of mode 2) as shown in Eqs. (14-18). It is possible to derive all the involved quantities for both $n = 2m$ and $n = 2m + 1$. The obtained results read

$$x_{+,2m,p}^{(l)} = (-1)^p \left(\frac{t}{r}\right)^p \sqrt{\frac{l!}{p!(l-p)!}}\frac{c_{2m}^{(l-p)}(\beta r)N_+^{(0)}(\beta t)}{c_{2m}^{(l)}(\beta r)N_+^{(p)}(\beta t)}; \quad 0 \le p \le l \tag{C3}$$

$$y_{+,2m,p}^{(l)} = (-1)^p \frac{t^{p-2}\sqrt{l!(l-p+1)!}\,c_{2m}^{(l+1-p)}(\beta r)N_-^{(0)}(\beta t)}{r^p(l-p)!\sqrt{(l+1)p!}\,c_{2m}^{(l+1)}(\beta r)N_-^{(p)}(\beta t)}\left(t^2 - \frac{p}{l-p+1}r^2\right); 0 \le p \le l \tag{C4}$$

$$y_{+,2m,p}^{(l)} = (-1)^l \frac{t^{l-1}c_{2m}^{(0)}(\beta r)N_-^{(0)}(\beta t)}{r^{l-1}c_{2m}^{(l+1)}(\beta r)N_-^{(l+1)}(\beta t)}; \quad p = l+1 \tag{C5}$$

$$B_{+,2m}^{(l)} = \frac{t\sqrt{(l+1)}c_{2m}^{(l+1)}(\beta r)N_+^{(0)}(\beta t)L_{+,2m}^{(l)}(\beta t)}{c_{2m}^{(l)}(\beta r)N_-^{(0)}(\beta t)K_{+,2m}^{(l)}(\beta t)}, \tag{C6}$$

$$P_{+,2m}^{(l)} = \frac{F^2(\beta r)|r|^{2l}\left|c_{2m}^{(l)}(\beta r)\right|^2 N_+^{(l)2}(\beta)}{N_+^{(0)2}(\beta t)L_{+,2m}^{(l)2}(\beta t)\mathfrak{N}_{+,2m}^{(l)2}}, \tag{C7}$$

for $n = 2m$, and

$$x_{+,2m,p}^{(l)} = (-1)^p \left(\frac{t}{r}\right)^p \sqrt{\frac{l!}{p!(l-p)!}}\frac{c_{2m+1}^{(l-p)}(\beta r)N_-^{(0)}(\beta t)}{c_{2m+1}^{(l)}(\beta r)N_-^{(p)}(\beta t)}; \quad 0 \le p \le l \tag{C8}$$

$$y_{+,2m+1,p}^{(l)} = (-1)^p \frac{t^{p-2}\sqrt{l!(l-p+1)!}\,c_{2m+1}^{(l+1-p)}(\beta r)N_+^{(0)}(\beta t)}{r^p(l-p)!\sqrt{(l+1)p!}\,c_{2m+1}^{(l+1)}(\beta r)N_+^{(p)}(\beta t)}\left(t^2 - \frac{p}{l-p+1}r^2\right); \quad 0 \le p \le l, \tag{C9}$$

$$y_{+,2m+1,p}^{(l)} = (-1)^l \frac{t^{l-1}c_{2m+1}^{(0)}(\beta r)N_+^{(0)}(\beta t)}{r^{l-1}c_{2m+1}^{(l+1)}(\beta r)N_+^{(l+1)}(\beta t)}, \tag{C10}$$

$$B_{+,2m+1}^{(l)} = \frac{t\sqrt{(l+1)}c_{2m+1}^{(l+1)}(\beta r)N_-^{(0)}(\beta t)L_{+,2m+1}^{(l)}(\beta t)}{c_{2m+1}^{(l)}(\beta r)N_+^{(0)}(\beta t)K_{+,2m+1}^{(l)}(\beta t)}, \tag{C11}$$

$$P_{+,2m+1}^{(l)} = \frac{F^2(\beta r)r^{2l}\left|c_{2m+1}^{(l)}(\beta r)\right|^2 N_+^{(l)2}(\beta)}{N_-^{(0)2}(\beta t)L_{+,2m+1}^{(l)2}(\beta t)\mathfrak{N}_{+,2m+1}^{(l)2}}, \tag{C12}$$

for $n = 2m + 1$.

Similarly, one can derive all the relevant quantities in the case of the general odd source state $|\Omega_-^{(l)}\rangle_1$ in Eq. (2). The difference will be only in some factors. Consider the difference on example of the quantities Eqs. (C3-C5). In the case of $|\Omega_-^{(l)}\rangle_1$ we must use the factor $N_-^{(0)}(\beta t)/N_-^{(p)}(\beta t)$ instead of $N_+^{(0)}(\beta t)/N_+^{(p)}(\beta t)$ in Eq. (C3) for $x_{-,2m,p}^{(l)}$. To obtain analytic expressions for $y_{-,2m,p}^{(l)}$ for $0 \le p \le l+1$ from Eqs. (C4,C5), we must use the substitution $N_-^{(0)}(\beta t)/N_-^{(p)}(\beta t) \to N_+^{(0)}(\beta t)/N_+^{(p)}(\beta t)$ in Eq. (C4) and

$N_-^{(0)}(\beta t)/N_-^{(l+1)}(\beta t) \to N_+^{(0)}(\beta t)/N_+^{(l+1)}(\beta t)$ in Eq. (C5). Such changes should be made also in Eqs. (C6,C7) in order to obtain analytical expressions for $B_{-,2m}^{(l)}$ and $P_{-,2m}^{(l)}$.

**Appendix D. Superimposing even/odd states (11,12) with delocalized photon**

To obtain analytical expressions for the amplitudes, it is worth making use of again the technique developed above. Consider it on example of input even CV state $|\Omega_+^{(01...l)}\rangle$ in Eq. (11) in the case of registration of even number $n = 2m$ photons in second auxiliary mode. Calculations give the following amplitudes

$$x_{+,2m,p}^{(01...l)} = (-1)^p \frac{t^p}{\sqrt{p!}} \frac{f_{+,2m,p}^{(01...l)} N_+^{(0)}(\beta t)}{f_{+,2m,0}^{(01...l)} N_+^{(p)}(\beta t)}, \tag{D1}$$

for the even CV state $|\Psi_{+,2m}^{(01...l)}\rangle = L_{+,2m}^{(01...l)} \sum_{p=0}^{l} x_{+,2m,p}^{(01...l)} |\Omega_+^{(p)}\rangle$ with $L_{+,2m}^{(01...l)}$ being the normalization factor, where new parameters are introduced

$$f_{+,2m,p}^{(01...l)} = \sum_{j=p}^{l} (-1)^j \frac{b_+^{(j)} N_+^{(j)}(\beta) r^{j-p} c_{2m}^{(j-p)}(\beta r) \sqrt{j!}}{\sqrt{(j-p)!}}, \tag{D2}$$

$$f_{+,2m,0}^{(01...l)} = \sum_{j=0}^{l} (-1)^j b_j^{(+)} N_+^{(j)}(\beta) r^j c_{2m}^{(j)} c_{j2m}(\beta r). \tag{D3}$$

The odd CV state is represented as $|\Phi_{+,2m}^{(01...l)}\rangle = K_{+,2m}^{(01...l)} \left( \sum_{p=0}^{k} y_{+,2m,p}^{(01...l)} |\Omega_-^{(p)}\rangle + \left( rtc_{2m}^0(\beta r) N_-^{(0)}(\beta t)/g_{+,2m,0}^{(01...l)} \right) \sum_{k=0}^{l} z_{+,2m,k}^{(01...l)} \Omega_-^{(k+1)} \right)$, where $K_{+,2m}^{(01...l)}$ is the normalization factor with amplitudes

$$y_{+,2m,p}^{(01...l)} = (-1)^p \frac{t^p}{\sqrt{p!}} \frac{g_{+,2m,p}^{(01...l)} N_-^{(0)}(\beta t)}{g_{+,2m,0}^{(01...l)} N_-^{(p)}(\beta t)}, \tag{D4}$$

$$z_{+,2m,k}^{(01...l)} = \frac{b_k^{(+)} N_+^{(k)}(\beta) t^k \sqrt{k+1}}{N_-^{(k+1)}(\beta t)}, \tag{D5}$$

$$g_{+,2m,p}^{(01...l)} = \sum_{j=p}^{l} (-1)^j \frac{b_j^{(+)} N_+^{(j)}(\beta) r^{j-p} c_{2m}^{(j-p+1)}(\beta r) \sqrt{j!(j-p+1)!}}{(j-p)!} \left( t^2 - \frac{p}{j-p+1} r^2 \right), \tag{D6}$$

$$g_{+,2m,0}^{(01...l)} = t^2 \sum_{j=0}^{l} (-1)^j b_j^{(+)} N_+^{(j)}(\beta) r^j c_{2m}^{(j+1)}(\beta r) \sqrt{j+1}. \tag{D7}$$

The parameter $B_{+,2m}^{(01...l)}$ largely determining the entanglement of the generated state becomes

$$B_{+,2m}^{(01...l)} = \frac{g_{+,2m,0}^{(01...l)} N_+^{(0)}(\beta t) L_{+,2m}^{(01...l)}}{t f_{+,2m,0}^{(01...l)} N_-^{(0)}(\beta t) K_{+,2m}^{(01...l)}}. \tag{D8}$$

The success probability to conditionally produce the hybrid entangled states is the following

$$P_{+,2m}^{(01...l)} = \frac{F^2(\beta r) |r|^{2l} \left| f_{+,2m,0}^{(01...l)} \right|^2 N_+^{(01...l)2}}{N_+^{(0)2}(\beta t) L_{+,2m}^{(01...l)2}(\beta t) \mathfrak{N}_{+,2m}^{(01...l)2}}, \tag{D9}$$

where $\mathfrak{N}_{+,2m}^{(01...l)}$ is the overall normalization factor of the conditional state.

It can be shown by direct calculations that the above expressions are transformed into those already introduced in the previous Appendixes B and C in the case of if all amplitudes of the input state $|\Omega_+^{(01...l)}\rangle$ in Eq. (11) take zero values $b_j^{(+)} = 0$ with the exception of one $b_l^{(+)} = 1$. The results can be extended to the case of recording an odd number of measurement outcomes $n = 2m + 1$. In the same way, the conditional hybrid entangled states can be analyzed in the case of using the input state $|\Omega_+^{(01...l)}\rangle$ in Eq. (19).

**Appendix E. Notes concerning generation of even/odd CV states**

Despite the fact that the problem of generating even odd *even/odd* states is beyond the scope of this work, here we will consider a method for generating such states in practice [41]. Example of the *even* state routinely generated in laboratories is single mode squeezed vacuum ($SMSV$) state $|SMSV\rangle = |even\rangle = \sum_{l=0}^{\infty} s_{2l}|2l\rangle$ with amplitudes

$$s_{2l} = \frac{(\tanh r)^l}{\sqrt{\cosh r}} \frac{\sqrt{(2l)!}}{2^l l!}, \tag{E1}$$

defined through the squeezing parameter $r$. This state can become the basis for creating other even/odd CV states.

Indeed, consider the passage the $SMSV$ state located in first mode (second mode in vacuum state) through the $BS$. After passing the $SMSV$ through the beam splitter $BS_{12}(|SMSV\rangle_1|0\rangle_2)$, the second mode of the output state is measured by $PNR$ detector. Depending on the measurement outcome in the second auxiliary mode, the conditional state is generated. The states can be derived using relation (C1). It is possible to show that if the measurement outcome of the $PNR$ detection in second auxiliary mode is even $2m$, then following conditional state

$$|\Upsilon_{2m}\rangle = L_{2m} \sum_{k=0}^{\infty} s_{2(k+m)} t^{2k} \sqrt{\frac{(2(m+k))!}{(2k)!}} |2k\rangle, \tag{E2}$$

is created, where $L_{2m}$ is the normalization factor. The conditional state is a superposition of exclusively even Fock states, therefore, it can be recognized as even CV. Suppose that odd measurement outcome $2m+1$ is registered in the second auxiliary mode, then next conditional state is generated

$$|\Upsilon_{2m+1}\rangle = L_{2m+1} \sum_{k=0}^{\infty} s_{2(k+m+1)} t^{2k} \sqrt{\frac{(2(m+k+1))!}{(2k+1)!}} |2k+1\rangle, \tag{E3}$$

where $L_{2m+1}$ is the normalization factor. Since the state already contains exclusively odd Fock states, it is odd CV state. Here, the subscripts $2m$ and $2m+1$ refer to the number of registered photons.

As shown, the states $|\Omega_{\pm}^{(l)}\rangle$ are not orthogonal to each other (Eq. (8)), which complicates the possibility of decomposing an arbitrary state in a given basis. Let us present a practical procedure that could be used to realize the states $|\Upsilon_{2m}\rangle$ and $|\Upsilon_{2m+1}\rangle$ in terms of superposition of $SDlPSs$ as given in Eqs. (11,12). Show it on example of the state $|\Omega_{+}^{(01...l)}\rangle$ in Eq. (11). For this purpose, let us represent $SDlPSs$ as a superposition of even Fock states $|\Omega_{+}^{(l)}\rangle = \sum_{n=0}^{\infty} g_{+}^{(ln)}|2n\rangle$, where the amplitudes $g_{+}^{(ln)}$ directly stem from Eq. (5), while the state $|\Upsilon_{2m}\rangle$ is rewritten as $|\Upsilon_{2m}\rangle = \sum_{n=0}^{\infty} f_{2n}^{(2m)}|2n\rangle$ whose amplitudes follows from (E2). Equating the amplitudes for the same even Fock states for the states $|\Omega_{+}^{(01...l)}\rangle$ and $|\Upsilon_{2m}\rangle$, one obtains a system of $l+1$ linear equations for $l+1$ unknown amplitudes $\{b_{+}^{(k)}; k=0,1,...,l\}$

$$\sum_{k=0}^{l} b_{+}^{(k)} g_{+}^{(kn)} = f_{2n}^{(2m)}. \tag{E4}$$

The solution of this system will make it possible to obtain the state $|\Omega_{+}^{(01...l)}\rangle$, especially with rather small amplitudes $\beta$, with a very high fidelity coinciding with $|\Upsilon_{2m}\rangle$. The same procedure can be applied to the state $|\Upsilon_{2m+1}\rangle$. This method of realizing even/odd CV states deserves a separate study.

**References**

[1] R. Horodecki, P. Horodecki, M. Horodecki, and K. Horodecki, "Quantum entanglement", Rev. Mod. Phys. **81**, 856-942 (2009).

[2] D. Kurzyk, "Introduction to quantum Entanglement", Theoretical and Applied Informatics **24**, 135-150 (2012).
[3] O. Gühne and G. Tóth, "Entanglement detection", Physics Reports **474**, 1-75 (2009).
[4] E. Andersson, and P. Öhberg (editors), Quantum Information and Coherence, Springer International Publishing, Switzerland, (2014).
[5] C. Bennett *et al.*, "Teleporting an unknown state via dual classical and Einstein-Podolsky-Rosen channels", Phys. Rev. Lett. **70**, 1895-1899 (1993).
[6] L. Vaidman, "Teleportation of quantum states", Phys. Rev. A. **49**, 1473-1476 (1994).
[7] D. Bouwmeester D *et al.*, "Experimental quantum teleportation", Nature **390**, 575-579 (1997).
[8] N. Lee *et al.*, "Teleportation of nonclassical wave packets of light", Science **332**, 330-333 (2011).
[9] S.L. Braunstein and H.J. Kimble, "Teleportation of continuous quantum variables", Phys. Rev. Lett. **80**, 869-872 (1998).
[10] S.A. Podoshvedov, "Quantum teleportation protocol with an assistant who prepares amplitude modulated unknown qubit", JOSA B. **35**, 861-877 (2018).
[11] S.A. Podoshvedov, "Efficient quantum teleportation of unknown qubit based on DV-CV interaction mechanism", Entropy **21**, 150 (2019).
[12] E.V. Mikheev, A.S. Pugin, D.A. Kuts, S.A. Podoshvedov, and N.B. An "Efficient production of large-size optical Schrödinger cat states", Scientific Reports **9**, 14301 (2019).
[13] O.S. Magaña-Loaiza *et al.*, "Multiphoton quantum-state engineering using conditional measurements", npj Quantum Information **5**, 80 (2019).
[14] D. Gottesman, and I.L. Chuang, "Demonstrating the viability of universal quantum computation using teleportation and single-qubit operations", Nature **402**, 390-393 (1999).
[15 R. Rausendorf, and H.J. Briegel, "A one-way quantum computer", Phys. Rev. Lett. **86**, 5188-5191 (2011).
[16] M.A. Nielsen, and I.L. Chuang, "Quantum Computation and Quantum Information", New York, Cambridge University Press, (2010).
[17] F. Arute *et al.*, "Quantum supremacy using a programmable superconducting processor", Nature **574**, 505-510 (2019).
[18] P. Kwiat, K. Mattle, H. Weinfurter, A. Zeilinger, A. Sergienko, and Y. Shih, "New high-intensity source of polarization-entangled photon pairs", Phys. Rev. Lett. **75**, 4337-4340 (1995).
[19] I. Schwartz, *et al.* "Deterministic generation of a cluster state of entangled photons", Science **354**, 434–437 (2016).
[20] S. Takeda, K. Takase, and A. Furusawa, "On-demand photonic entanglement synthesizer", Sci. Adv. **5**, eaaw4530 (2019).
[21] A. I. Lvovsky, R. Ghobadi, A. Chandra, A. S. Prasad, and C. Simon, "Observation of micro-macro entanglement of light", Nature Phys. **9**, 541-544 (2013).
[22] K. Huang, H. Le Jeannic, O. Morin, T. Darras, G. Guccione, A. Cavailles, J. Laurat, "Engineering optical hybrid entanglement between discrete- and continuous-variable states", New J. Phys. **21**, 083033 (2019).
[23] S.A. Podoshvedov and B.A. Nguyen, "Designs of interactions between discrete- and continuous-variable states for generation of hybrid entanglement" Quantum Inf. Process. **18**, 68 (2019).
[24] K. Huang, H.L. Jeannic, O. Morin, T. Darras, G. Guccione, A. Cavaillès, and J. Laurat, "Engineering optical hybrid entanglement between discrete- and continuous-variable states", New J. Phys. **21**, 083033 (2019).
[25] H. Jeong, A. Zavatta, M. Kang, S.-W. Lee, L.S. Costanzo, S. Grandi, T.C. Ralph, and M. Bellini, "Generation of hybrid entanglement of light", Nat. Photonics **8**, 564–569 (2014).

[26] E. Agudelo, J. Sperling, L.S. Costanzo, M. Bellini, A. Zavatta, and W. Vogel, "Conditional hybrid nonclassicality", Phys. Rev. Lett. **119**, 120403 (2017).
[27] A.S. Rab, E. Polino, Z.-X. Man, N. Ba An, Y.-J. Xia, N. Spagnolo, R. Lo Franco, and F. Sciarrino, "Entanglement of photons in their dual wave-particle nature", Nature Commun. **8**, 915 (2017).
[28] N. Biagi, L.S. Costanzo, M. Bellini, and A. Zavatta, "Entangling macroscopic light states be delocalized photon addition" Phys. Rev. Lett. **124**, 033604 (2020).
[29] A.E. Ulanov, D. Sychev, A.A. Pushkina, I.A. Fedorov, and A.I. Lvovsky, "Quantum teleportation between discrete and continuous encodings of an optical qubit", Phys. Rev. Lett. **118**, 160501 (2017).
[30] D.V. Sychev, A.E. Ulanov, E.S. Tiunov, A.A. Pushkina, A. Kuzhamuratov, V. Novikov, and A.I. Lvovsky, "Entanglement and teleportation between polarization and wave-like encodings of an optical qubit", Nat. Comm. **9**, 3672 (2018).
[31] A. Peres, "Separability criterion for density matrices", Phys. Rev. Lett. **77**, 1413-1415 (1996).
[32] G. Vidal, and R.F. Werner, "Computable measure of entanglement", Phys. Rev. A. **65**, 032314, (2002).
[33] B. Yurke and D. Stoler, "Generating quantum mechanical superpositions of macroscopically distinguishable states via amplitude dispersion," Phys. Rev. Lett. **57**, 13-16 (1986).
[34] S.J. van Enk and O. Hirota, "Entangled coherent states: teleportation and decoherence," Phys. Rev. A **64**, 022313 (2001).
[35] S.A. Podoshvedov, "Building of one-way Hadamard gate for squeezed coherent states", Phys. Rev. A. **87**, 012307 (2013).
[36] S.A. Podoshvedov, "Generation of displaced squeezed superpositions of coherent states", J. Exp. Theor. Phys. **114**, 451-464 (2012).
[37] S.A. Podoshvedov, "Elementary quantum gates in different bases", Quant. Information Processing **15**, 3967–3993 (2016).
[38] D.N. Klyshko, "The nonclassical light", Physics-Uspekhi **166**, 613-638 (1996).
[39] R.J. Birrittella, M.E. Baz, and C.C. Gerry "Photon catalysis and quantum state engineering," J Opt. Soc. Am. B **35**, 1514-1524 (2018).
[40] A.E. Lita, A. J. Miller and S. W. Nam, "Counting near-infrared single-photons with 95% efficiency," Optics Express **16**, 3032-3040 (2008).
[41] M. Dakna, T. Anhut, T. Opatrny, L. Knöll, and D. G. Welsch, "Generating Schrödinger-cat-like state by means of conditional measurement on a beam splitter," Phys. Rev. A **55**, 3184-3194 (1997).

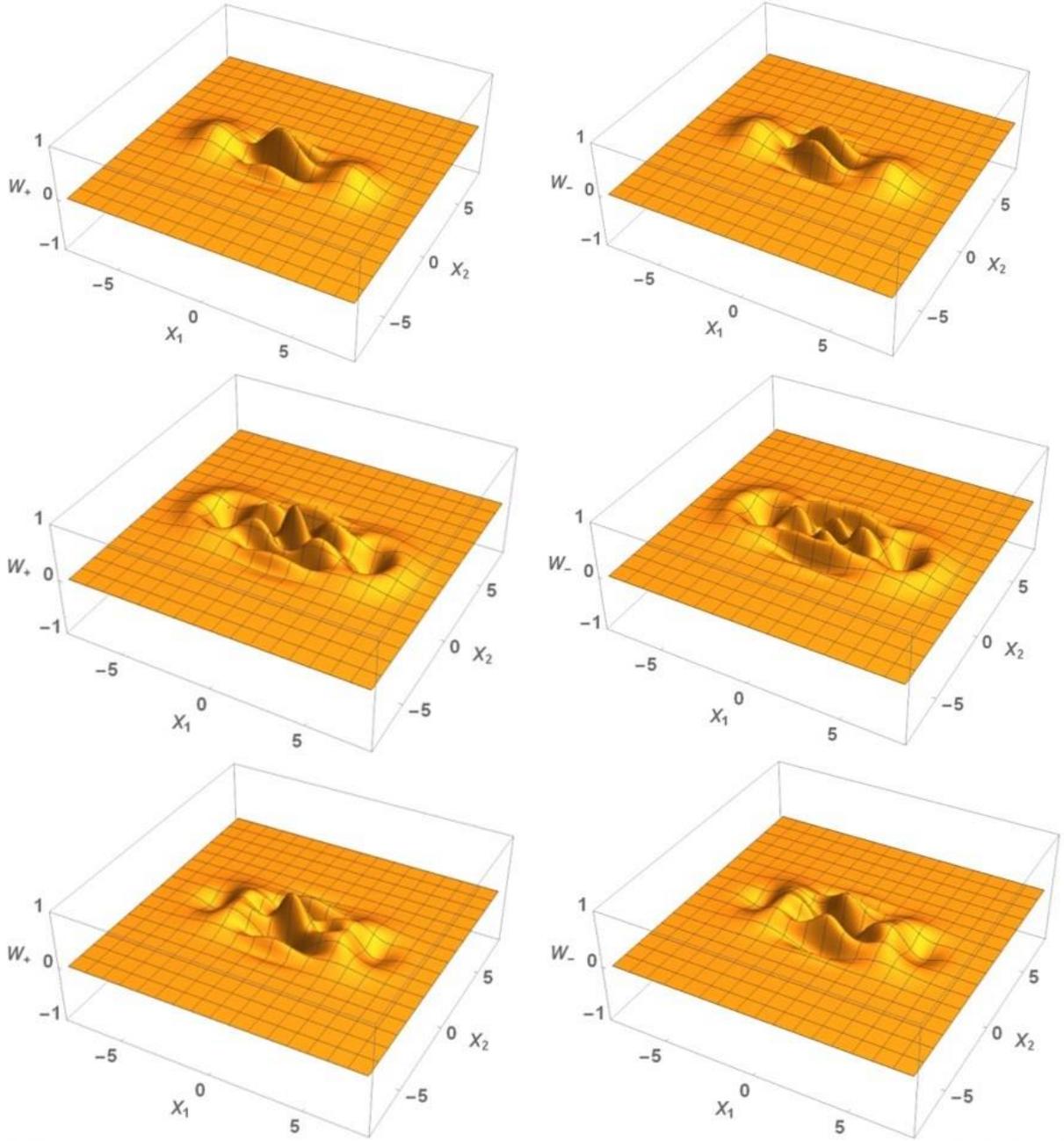

**FIG. 1.** Wigner functions $W$ for three types of states: $|\Omega_\pm^{(0)}\rangle$ (top row), $|\Omega_\pm^{(1)}\rangle$ (middle row) and $|\Omega_\pm^{(01)}\rangle$ (bottom row) for $\beta = 2$.

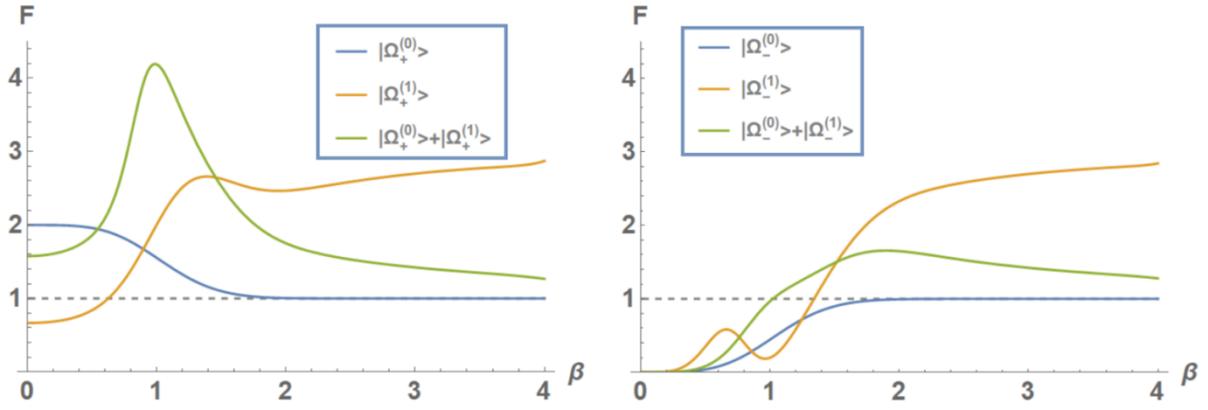

**Fig. 3.** Dependence of the Fano factor $F$ on the size $\beta$ of the states $|\Omega_\pm^{(0)}\rangle$, $|\Omega_\pm^{(1)}\rangle$ and $|\Omega_\pm^{(01)}\rangle$.

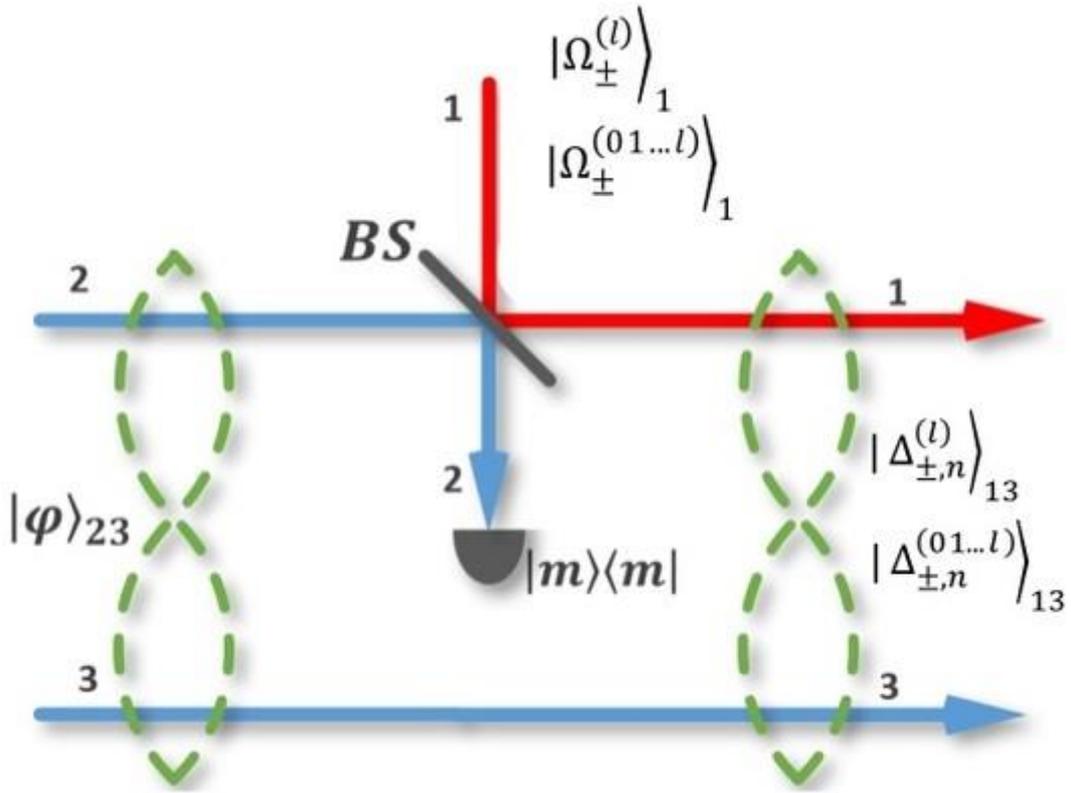

**FIG. 4.** Scheme for generation of macro-micro hybrid entangled light using the CV states $|\Omega_\pm^{(l)}\rangle$ in Eqs. (5,6) and $|\Omega_\pm^{(01\ldots l)}\rangle$ in Eqs. (11,12) as sources together with the delocalized photon state $|\varphi\rangle_{23}$ in Eq. (13). Heralded entanglement $|\Delta_m^{(l\pm)}\rangle_{13}$ with some negativity either $\mathcal{N}_{2m}^{(l\pm)}$ or $\mathcal{N}_{2m+1}^{(l\pm)}$ occurs every time a measurement $m$ ($m$ can be either even or odd) is recorded in the second auxiliary mode. Under certain experimental conditions $(\beta, t)$, the entanglement can take on the maximum possible value $\mathcal{N}_{max} = 1$. Truncated versions of the *SDlPSs* in Eqs. (9,10) can also be used to generate the entangled states.

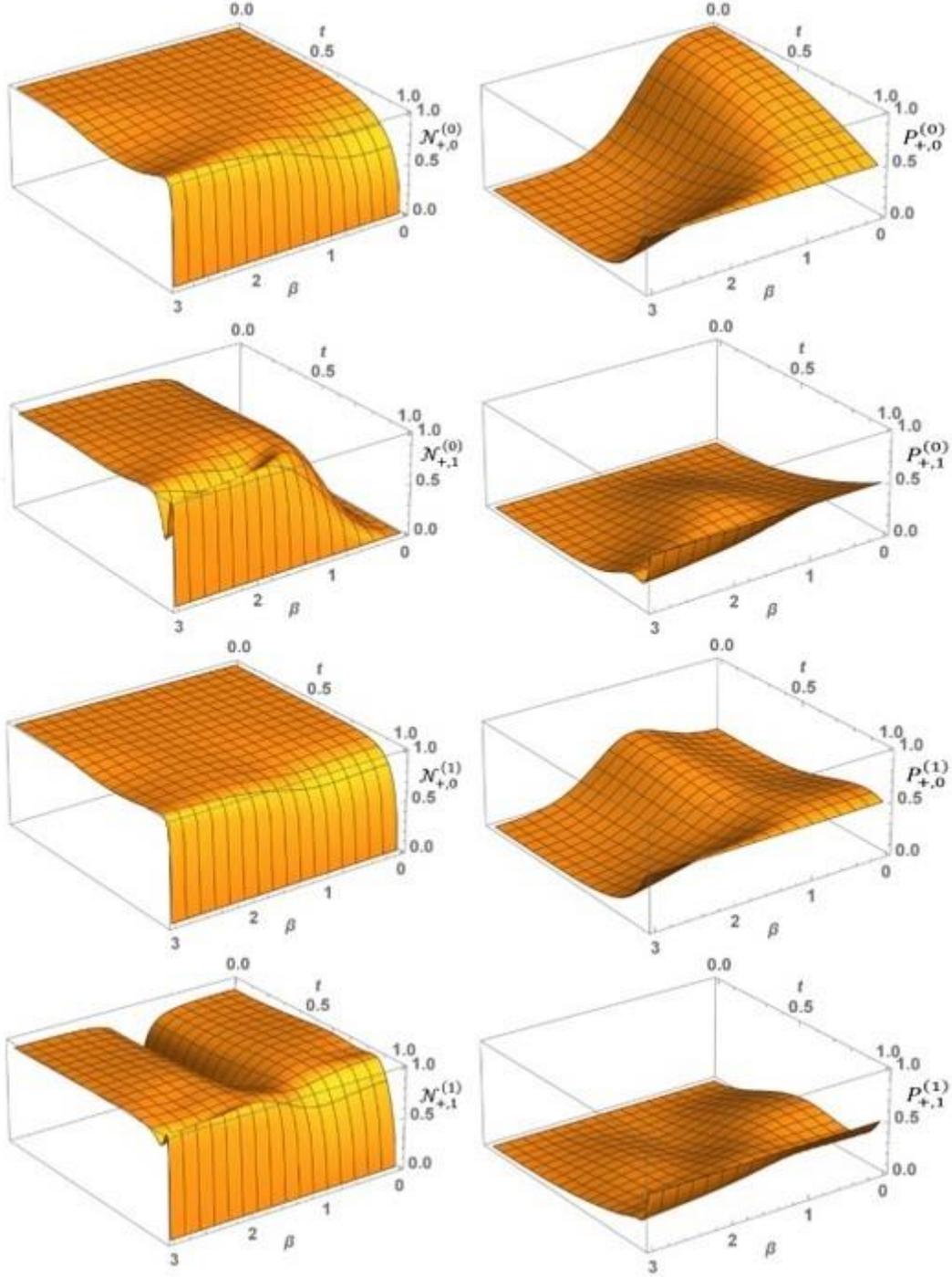

**FIG. 5.** Plots of the negativities $\mathcal{N}_{+,0}^{(0)}, \mathcal{N}_{+,1}^{(0)}, \mathcal{N}_{+,0}^{(1)}$ and $\mathcal{N}_{+,1}^{(1)}$ (left column) as well as the probabilities $P_{+,0}^{(0)}, P_{+,1}^{(0)}, P_{+,0}^{(1)}$ and $P_{+,1}^{(1)}$ (right column) in dependency on $\beta$ and $t$.

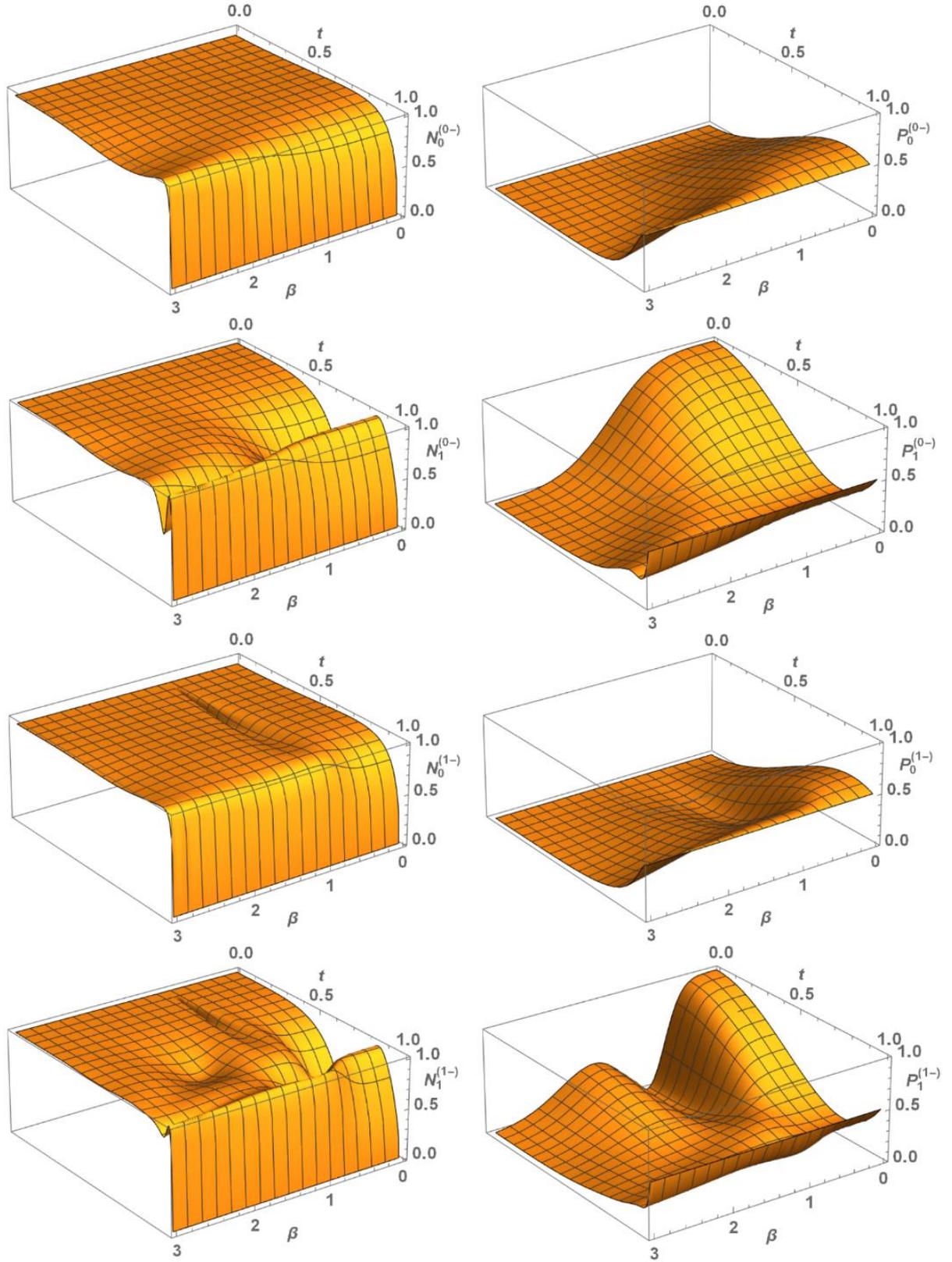

**FIG. 6.** Plots of the negativities $\mathcal{N}^{(0)}_{-,0}, \mathcal{N}^{(0)}_{-,1}, \mathcal{N}^{(1)}_{-,0}$ and $\mathcal{N}^{(1)}_{-,1}$ (left column) as well as the probabilities $P^{(0)}_{-,0}, P^{(0)}_{-,1}, P^{(1)}_{-,0}$ and $P^{(1)}_{-,1}$ (right column) in dependency on $\beta$ and $t$.

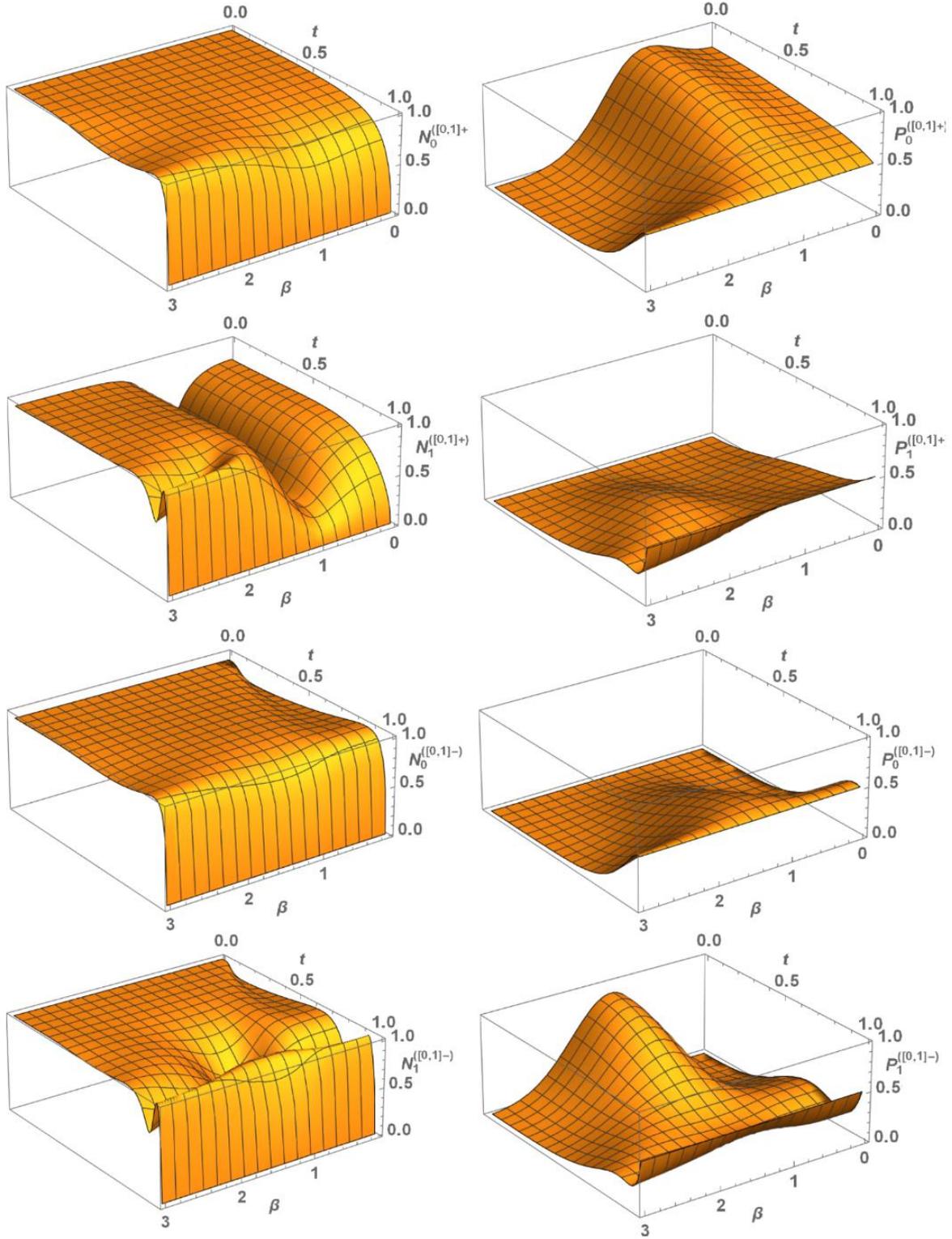

**FIG. 7.** Plots of the negativities $\left(\mathcal{N}_{+,0}^{(01)}, \mathcal{N}_{+,1}^{(01)}, \mathcal{N}_{-,0}^{(01)}, \mathcal{N}_{-,1}^{(01)}\right)$ (left column) as well as the probabilities $P_{+,0}^{(01)}, P_{+,1}^{(01)}, P_{-,0}^{(01)}$ and $P_{-,1}^{(01)}$ (right column) in dependency on $\beta$ and $t$.

22